\documentclass[12pt]{article}
\pdfoutput=1   %
\usepackage{graphicx}
\usepackage{multirow}
\usepackage{epstopdf}  	%
\usepackage{epsfig}     %
\usepackage{SIunits}
\def\Dkspp{D^0\rightarrow K_S^0\pi^+\pi^-}

\def\Dkshh{D^0\rightarrow K_S^0hh}
\def\DDbar{D^0-\overline{D}^0}
\def\pbnr{}
\def\speaker{Longke LI}
\def\onbehalfof{Belle Collaboration}
\def\title{$\DDbar$ Mixing and CP Violation in $\Dkshh$ Measurements}
\def\affiliation{Department of Modern Physics\\
University of Science and Technology of China, Hefei, China}
\def\support{Supported by National Natural Science Foundation of China(No.10935008 and No.10875115)}

\newcommand\pubnumber{\pbnr}
\newcommand\pubdate{\today}
\def\Title#1{\begin{center} {\Large #1 } \end{center}}
\def\Author#1{\begin{center}{ \sc #1} \end{center}}

\newcommand{\OnBehalf}[1]{\sbox0{#1}\ifdim\wd0=0pt
        {}
	\else
	{\\on behalf of #1}
	\fi}
\newcommand{\SupportedBy}[1]{\sbox0{#1}\ifdim\wd0=0pt
        {}
	\else
	{\footnote{#1}}
	\fi}
\def\Address#1{\begin{center}{ \it #1} \end{center}}

\newcommand\pubblock{\includegraphics[width=5cm]{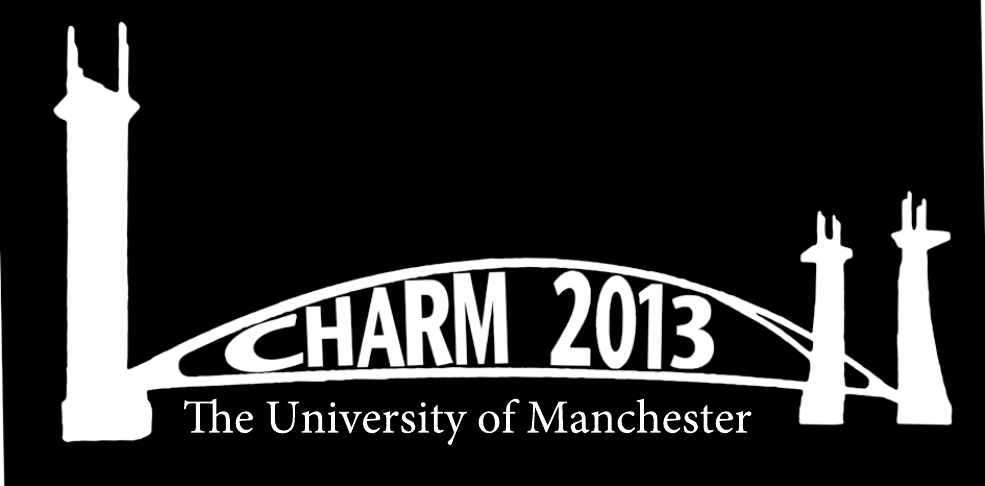}\hfill{\begin{tabular}{l} \pubnumber\\
         \pubdate  \end{tabular}}}
\newenvironment{Abstract}{\begin{quotation}  }{\end{quotation}}
\newenvironment{Presented}{\begin{quotation} \begin{center} 
             PRESENTED AT\end{center}\bigskip 
      \begin{center}\begin{large}}{\end{large}\end{center} \end{quotation}}
\def\Acknowledgements{\bigskip  \bigskip \begin{center} \begin{large}
             \bf ACKNOWLEDGEMENTS \end{large}\end{center}}
\def\venue{The 6$^{th}$ International Workshop on Charm Physics\\
(CHARM 2013)\\
Manchester, UK,  31 August -- 4 September, 2013}

\def\beq{\begin{equation}}
\def\eeq#1{\label{#1}\end{equation}}
\def\eeqn{\end{equation}}

\def\beqa{\begin{eqnarray}}
\def\eeqa#1{\label{#1}\end{eqnarray}}
\def\eeqan{\end{eqnarray}}

\let\bar=\overbar

\def\Dslash{\not{\hbox{\kern-4pt $D$}}}
\def\dslash{\not{\hbox{\kern-2pt $\del$}}}

\def\msb{{\bar{\ssstyle M \kern -1pt S}}}

\begin{document}
\begin{titlepage}
\pubblock

\vfill
\Title{\title}
\vfill
\Author{\speaker\SupportedBy{\support}\OnBehalf{\onbehalfof}}
\Address{\affiliation}
\vfill
\begin{Abstract}
In these proceedings, we give a summary of the experimental results from
CLEO, Belle, BaBar and CDF collaboration about $\DDbar$ mixing and CP violation
in self-conjugated-three-body decays $\Dkshh$ (where $h$ can be
$\pi$ or $K$).
We report preliminary results of measurement of $\DDbar$ mixing and indirect
CP violation in $\Dkspp$ decays using full data sample collected
by Belle detector at KEKB asymmetric-energy $e^+e^-$ collider:
give values for $x=(0.56\pm0.19^{+0.03+0.06}_{-0.09-0.09})\%$,
$y=(0.30\pm0.15^{+0.04+0.03}_{-0.05-0.06})\%$,
$|q/p|=0.90^{+0.16+0.05+0.06}_{-0.15-0.04-0.05}$
and $\arg(q/p)=-6\pm11^{+3+3}_{-3-4}(\degree)$.
We also report results of recent measurement searching for CP violation in
$\Dkspp$ decays by CDF collaboration using $6.0~fb^{-1}$ of data collected in
$p\bar{p}$ collisions at Tevatron. The phase-space-integrated CP asymmetry
is measured to be $A_{CP}=(-0.05\pm0.57\pm0.54)\%$ and the CP symmetry is also found to
be conserved in all individual intermediate contributions.
\end{Abstract}
\vfill
\begin{Presented}
\venue
\end{Presented}
\vfill
\end{titlepage}
\def\thefootnote{\fnsymbol{footnote}}
\setcounter{footnote}{0}
%

\section{Introduction and Mixing formalism}
The phenomenon of $\DDbar$ mixing has been of great interest
since Belle collaboration at KEK and BaBar collaboration at SLAC reported evidence
for it in 2007~\cite{evidenceBelle,evidenceBaBar}. This year LHCb collaboration
reported first observation of the $\DDbar$ oscillations from a single measurement
with significance corresponding to 9.1 standard deviations~\cite{observeLHCb}.

The two mass eigenstates $D_1$ and $D_2$, different to flavor eigenstates, are given by
\begin{eqnarray}
|D_{1,2}\rangle=p|D^0\rangle\pm q|\overline{D}^0\rangle  \label{eqn:mixingreason}
\end{eqnarray}
where $|p|^2+|q|^2=1$ holds and CPT invariance has been assumed. The mixing parameters
$x$ and $y$ in neutral $D$ meson system are defined as
\begin{eqnarray}
x\equiv\frac{M_2-M_1}{\Gamma},~y\equiv\frac{\Gamma_2-\Gamma_1}{2\Gamma}. \label{eqn:mixingpar}
\end{eqnarray}
where $M_{1,2}$ and $\Gamma_{1,2}$ are
the mass and width of $D_{1,2}$ and $\Gamma=(\Gamma_1+\Gamma_2)/2$.

The proper time evolution of the mass eigenstates is $|D_{1,2}(t)\rangle=e_{1,2}(t)|D_{1,2}\rangle$,
where $e_{1,2}(t)=e^{-i(M_{1,2}-(i\Gamma_{1,2}/2))t}$.
A state, which is prepared as a flavor eigenstate $|D^0\rangle$ or
$|\overline{D}^0\rangle$ at $t=0$, will evolve according to
\begin{eqnarray}
|D^0(t)\rangle&=&\frac{1}{2p}[p(e_1(t)+e_2(t))|D^0\rangle+q(e_1(t)-e_2(t))|\overline{D}^0\rangle]\nonumber,\\
|\overline{D}^0(t)\rangle&=&\frac{1}{2q}[p(e_1(t)-e_2(t))|D^0\rangle+q(e_1(t)+e_2(t))|\overline{D}^0\rangle].
\end{eqnarray}

The decay amplitude for three-body $D^0(\overline{D}^0)\rightarrow K_S^0h^+h^-$ decays,
${\cal{A}}({\cal\overline{A}})(m^2_+,m^2_-)$, depends on two kinematic variables:
$m^2_+=m_{K_S^0\pi^+}$ and $m^2_-=m_{K_S^0\pi^-}$.
The time-dependent decay amplitude for initially produced $D^0$ or $\overline{D}^0$ meson is then given by
\begin{eqnarray}
{\cal{M}}(m^2_+,m^2_-,t)&=&{\cal A}(m_+^2,m_-^2)\frac{e_1(t)+e_2(t)}{2}+\frac{q}{p}\overline{\cal A}(m_-^2,m_+^2)\frac{e_1(t)-e_2(t)}{2} \nonumber,\\
\overline{\cal{M}}(m^2_+,m^2_-,t)&=&\overline{\cal A}(m_+^2,m_-^2)\frac{e_1(t)+e_2(t)}{2}+\frac{p}{q}{\cal A}(m_-^2,m_+^2)\frac{e_1(t)-e_2(t)}{2}.
\label{eqn:timedep}
\end{eqnarray}

The decay rates as function of time are given by squaring the time-dependent amplitudes:
\begin{eqnarray}
|{\cal{M}}|^2&=&|e_1(t)|^2|A_1|^2+|e_2(t)|^2|A_2|^2+2{\cal{R}}[e_1(t)e_2^*(t)A_1A_2^*] \nonumber\\
&=&\left\{|A_1|^2e^{-yt}+|A_2|^2e^{yt}+2{\cal{R}}[A_1A_2^*]\cos(xt)+2{\cal{I}}[A_1A_2^*]\sin(xt)\right\}e^{-t}\nonumber, \\
|\overline{\cal{M}}|^2&=&\left\{|\overline{A}_1|^2e^{-yt}+|\overline{A}_2|^2e^{yt}+2{\cal{R}}[\overline{A}_1\overline{A}_2^*]\cos(xt)
+2{\cal{I}}[\overline{A}_1\overline{A}_2^*]\sin(xt)\right\}e^{-t}. \label{eqn:decayrate}
\end{eqnarray}
Here $t$ is in unit of $D^0$ lifetime. $y$ modifies the lifetime of
certain contributions to the Dalitz plot while $x$ introduces a
sinusoidal rate variation.

\section{Measurement Summary}
The $D^0\rightarrow K_S^0hh$
measurements from all experiments are shown in the Table~\ref{tab:d02kshh}.
The Heavy Flavor Averaging Group gives $D^0-\overline{D}^0$ Dalitz plot results assuming
no CP violation: $x=(0.419\pm0.211)\%$ and $y=(0.456\pm0.186)\%$~\cite{WA}.
\begin{table}[!hbtp]
  \begin{center}
    \begin{tabular}{c|c|c|c|c}
      \hline \hline
      Exp.          & Year                  & Data                          &      Channel              & Results            \\ \hline
      \multirow{2}{*}{CLEO} & \multirow{2}{*}{2005\cite{CLEO2005}} & \multirow{2}{*}{$9.0fb^{-1}$}      & \multirow{2}{*}{$h=\pi$} & $x=(1.8^{+3.4}_{-3.2}\pm0.4\pm0.4)\%$   \\
                    &                       &                               &                           & $y=(-1.4^{+2.5}_{-2.4}\pm0.8\pm0.4)\%$   \\\hline
      \multirow{5}{*}{Belle}  & \multirow{4}{*}{2007\cite{Belle2007}} & \multirow{4}{*}{$540fb^{-1}$}   & \multirow{4}{*}{$h=\pi$} & $x=(0.80\pm0.29^{+0.09+0.10}_{-0.07-0.14})\%$   \\
                    &                       &                               &                           & $y=(0.33\pm0.24^{+0.08+0.06}_{-0.12-0.08})\%$  \\ \cline{5-5}
                    &                       &                               &                           & $|q/p|=0.95^{+0.22+0.10}_{-0.20-0.09}$   \\
                    &                       &                               &                           & $\arg(q/p)=-0.035^{+0.17}_{-0.19}\pm0.09$  \\ \cline{2-5}
                    & 2009\cite{Belle2009}  & $673fb^{-1}$                  & $h=K$                     & $y_{CP}=(+0.11\pm0.61\pm0.52)\%$    \\ \hline
      \multirow{6}{*}{BaBar}  & \multirow{6}{*}{2010\cite{BaBar2010}} & \multirow{6}{*}{$469fb^{-1}$}   & \multirow{2}{*}{$h=\pi/K$} &    $x=(0.16\pm0.23\pm0.12\pm0.08)\%$  \\
                    &                       &                               &                           & $y=(0.57\pm0.20\pm0.13\pm0.07)\%$  \\\cline{4-5}
                    &                       &                               & \multirow{2}{*}{$h=\pi$}  & $x=(+0.26\pm0.24)\%$    \\
                    &                       &                               &                           &   $y=(0.60\pm0.21)\%$  \\\cline{4-5}
                    &                       &                               & \multirow{2}{*}{$h=K$}    &    $x=(-1.36\pm0.21)\%$    \\
                    &                       &                               &                           &   $y=(0.44\pm0.57)\%$   \\\hline
      CDF           & 2012\cite{CDF2012}    & $6.0fb^{-1}$                  & $h=\pi$                   & $A_{CP}=(-0.05\pm0.57\pm0.54)\%$    \\
      \hline  \hline
    \end{tabular}
    \caption{Self-conjugated decay $D^0\rightarrow K_{S}hh$ (here $h=K$ or $\pi$) published measurements from all experiments.}
    \label{tab:d02kshh}
  \end{center}
\end{table}

\section{CP violation asymmetries measuremt at CDF}
In the analysis of time-integrated CP violation asymmetries $A_{CP}$ in $D^0/\overline{D}^0\rightarrow K_S^0\pi^+\pi^-$
from CDF~\cite{CDF2012}, they exploit a large sample of $D^*(2012)^{\pm}$ decays using CDF II data with $6.0~fb^{-1}$ of
integrated luminosity produced in $p\bar{p}$ collision at $\sqrt{s}=1.96TeV$.
The neutral $D$ meson production flavor is determined by the charge of the pion in the $D^{*+}(2010)\rightarrow D^0\pi^+$
and $D^{*-}(2010)\rightarrow \overline{D}^0\pi^-$ decay ($D^*$ tagging).

They use an artificial neural network to
distinguish signal and background. The network uses five input variables:
the transverse decay length of the $D^0$ candidate divided by its resolution
$L_{xy}/\sigma_{L_{xy}}$($D^0$), the $\chi^2$ quality of the $D^{*+}$ vertex fit,
the impact parameter of the pion from the $D^{*+}$ decay divided by its uncertainty
$d_0/\sigma_{d_0}$($\pi_{D^{*+}}$), the transverse momentum of pion from $D^{*+}$
decay $p_T$($\pi_{D^{*+}}$) and the reconstructed mass of the $K_S^0$ candidate.
The $D^{*+}$ network training is based on the distribution of mass difference $\Delta M=M(K_S^0\pi^+\pi^-\pi^0)
-M(K_S^0\pi^+\pi^-)$ in the range $140<\Delta M<156~MeV/c^2$. The final neural network output requirement is chosen to maximize
$S/\sqrt{S+B}$ where $S$($B$) is the estimated number of signal(background) events in the signal region estimated from a fit
to the $M$($K_S^0\pi^+\pi^-$) distribution. For the Dalitz plot studies, the analysis is restricted to candidates populating two mass range,
$1.84<M(K_S^0\pi^+\pi^-)<1.89~GeV/c^2$ and $143.4<\Delta M<147.4~MeV/c^2$.
The selected data sample contains approximately $3.5\times10^5$ signal events
and consists of about $90\%$ correctly $D^*$-tagged $D^0$ signal, $1\%$ mistagged $D^0$ signal, and $9\%$ background candidates.

\begin{figure}[!hbtp]
  \begin{center}
  \includegraphics[width=0.25\paperwidth]{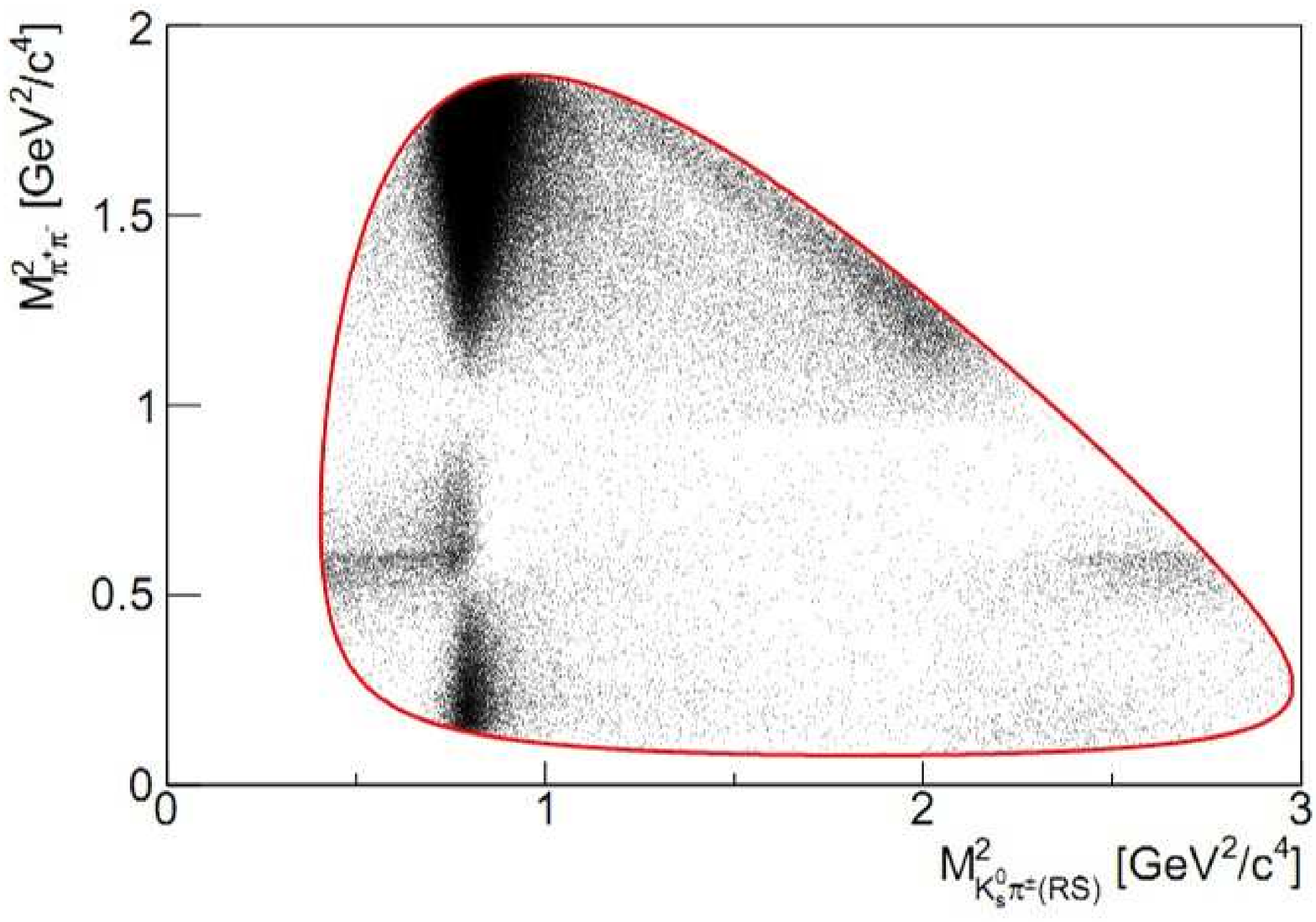}%
  \includegraphics[width=0.25\paperwidth]{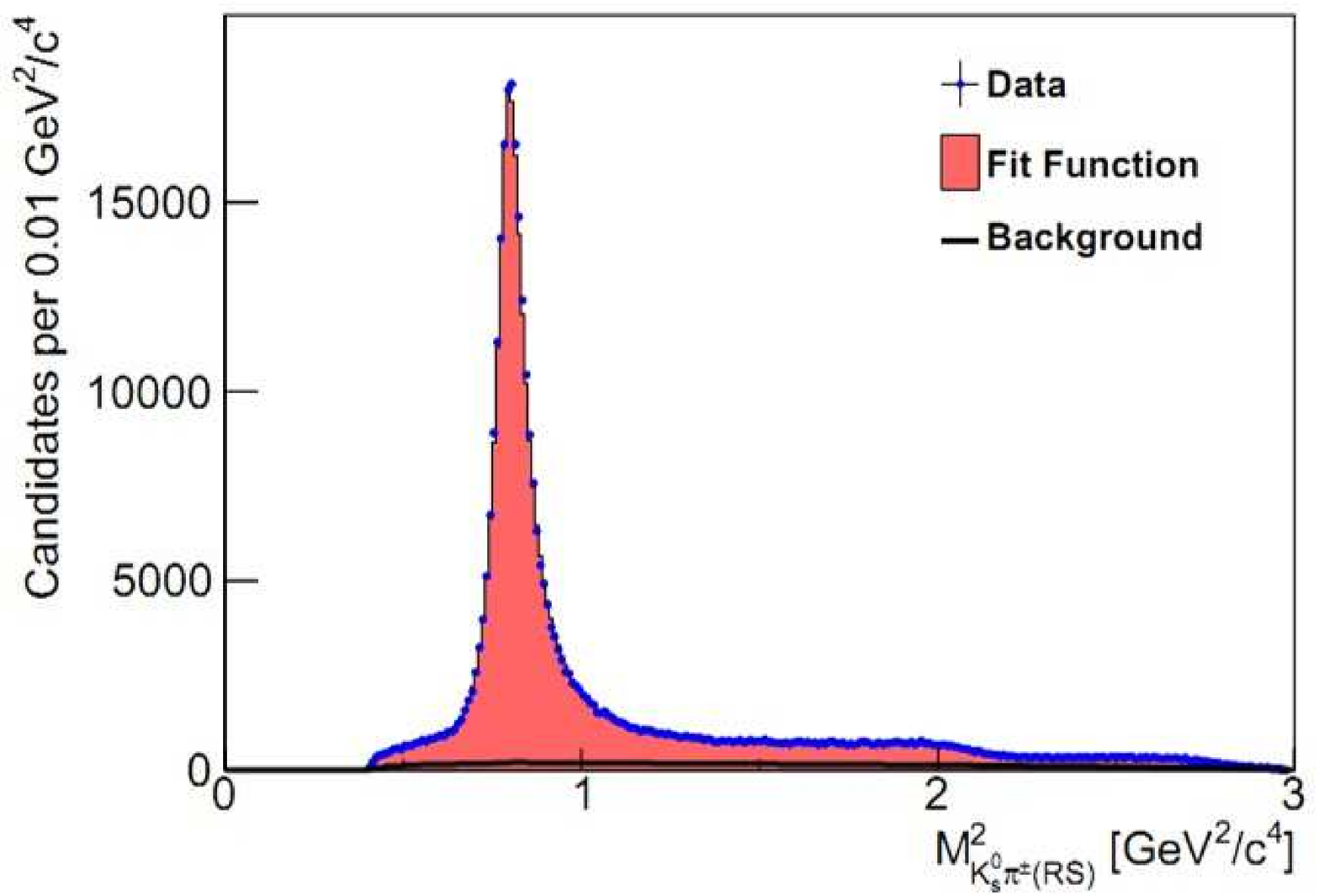}\\
  \includegraphics[width=0.25\paperwidth]{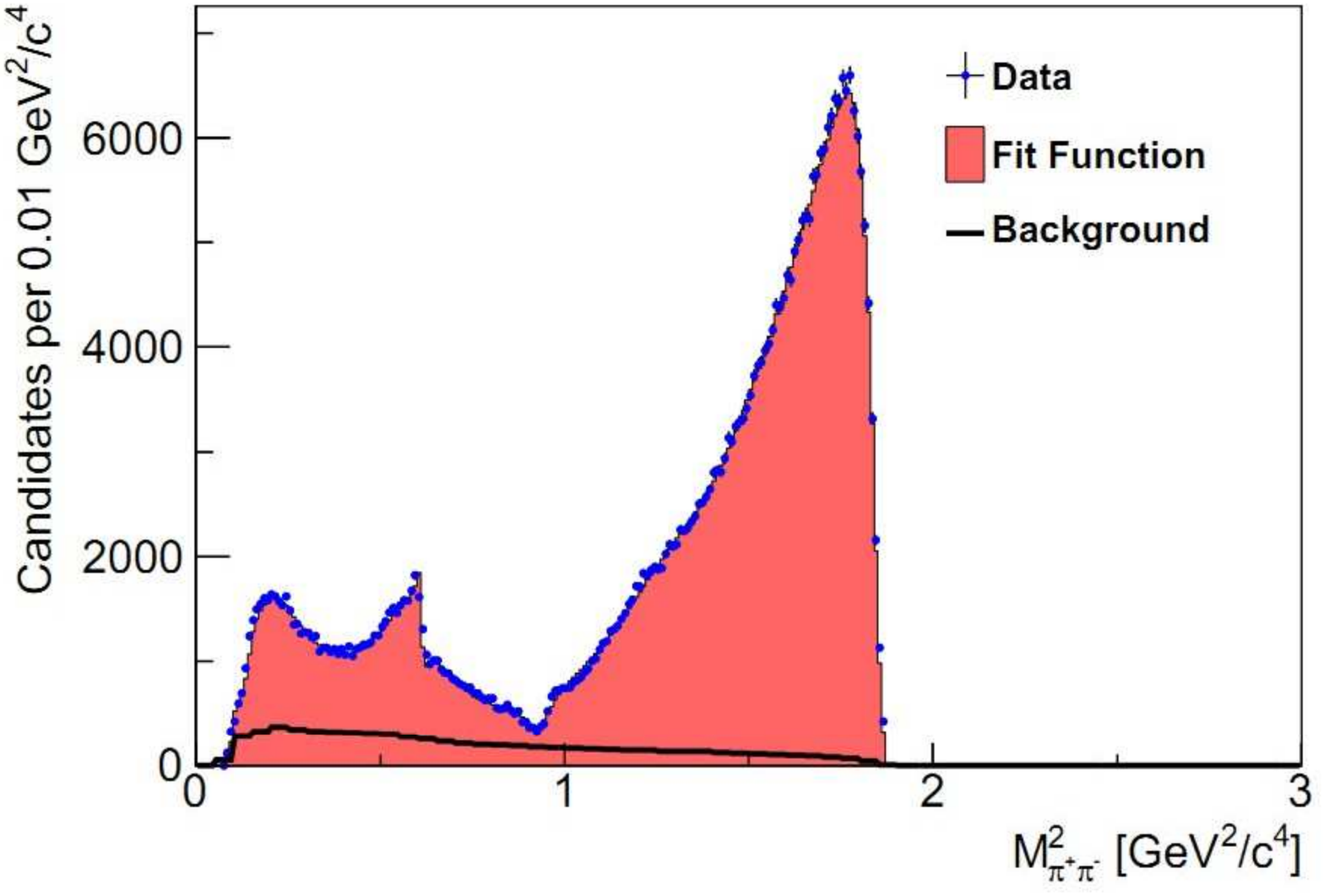}%
  \includegraphics[width=0.25\paperwidth]{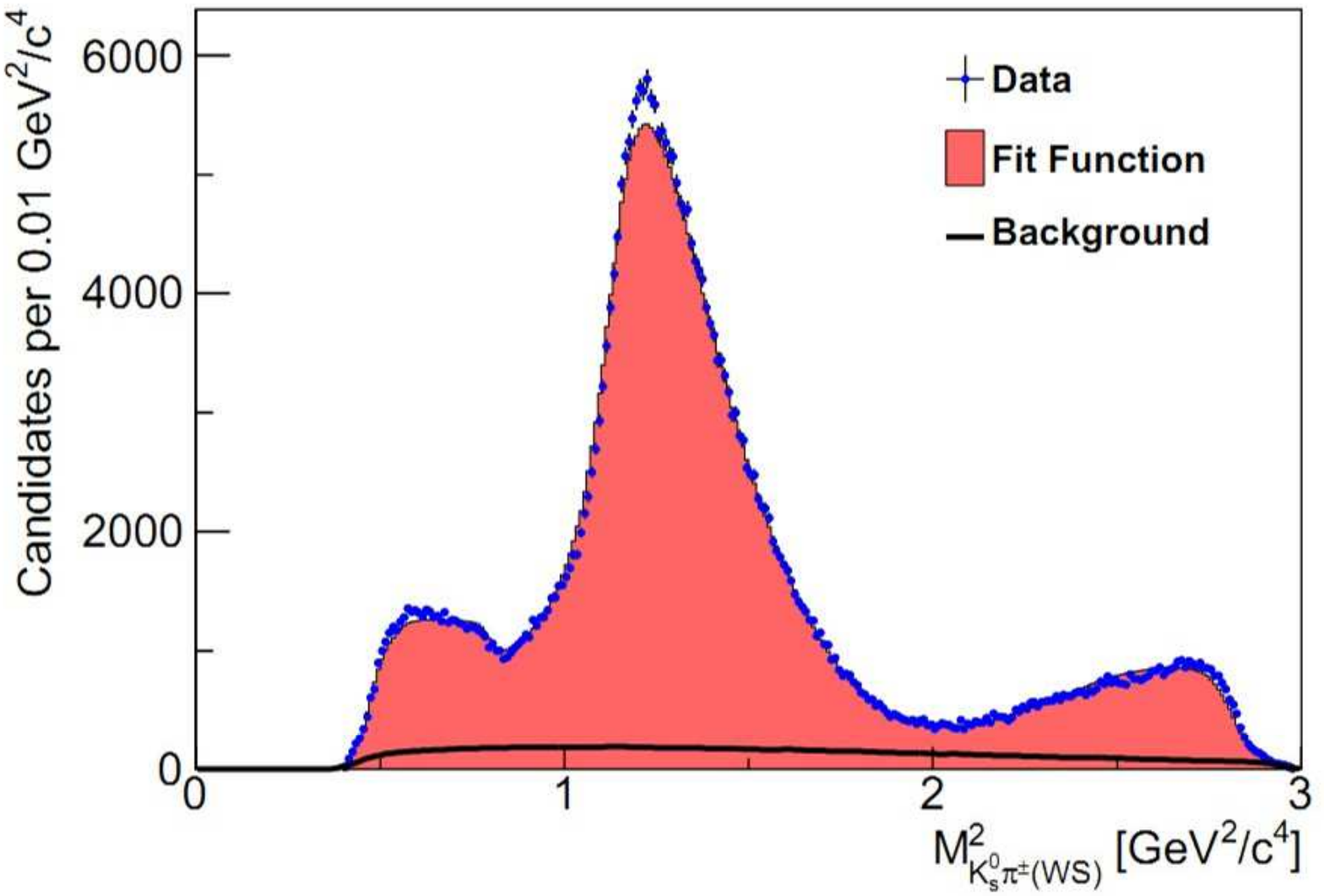}
  \vskip-10pt
  \caption{Dalitz plot of $D^0/\overline{D}^0\rightarrow K_S^0\pi^+\pi^-$ and projections of Dalitz-plot fit on the invidual two-body masses.}
  \label{fig:CDF_Dalitzplot}
  \end{center}
\end{figure}

The Dalitz plot of $D^0\rightarrow K_S^0\pi^+\pi^-$ decay, shown in Figure~\ref{fig:CDF_Dalitzplot},
contains three types of intermediate states contribution: Cabibbo-favored, doubly-Cabibbo-suppressed and CP eigenstates.
A measure for the overall integrated CP asymmetry is given by Eq.~\ref{eqn:intCPasym},
where ${\cal M}$ is the matrix element used in isobar model for $D^0$ decay and $\overline{\cal M}$
the one for $\overline{D}^0$ decay.
\begin{eqnarray}
A_{CP}=\frac{\int\frac{|{\cal M}|^2-|\overline{\cal M}|^2}{|{\cal M}|^2+|\overline{\cal M}|^2}dM^2_{K_S^0\pi^{\pm}(RS)}dM^2_{\pi^+\pi^-}}{\int dM^2_{K_S^0\pi^{\pm}(RS)}dM^2_{\pi^+\pi^-}}  \label{eqn:intCPasym}
\end{eqnarray}
\begin{eqnarray}
{\cal M}(\overline{\cal M})=a_{0}e^{i\delta_0}+\sum_{j}a_{j}e^{i(\delta_j\pm\phi_j)}(1\pm\frac{b_j}{a_j}){\cal A}_j.
\label{eqn:elematrix}
\end{eqnarray}

All CP violation quantities are found to be consistent with zero. The results for the CP violation amplitudes and phase,
defined in Eq.~\ref{eqn:elematrix} and obtained from simutaneous fit to the $D^0$ and $\overline{D}^0$ Dalitz
plots, are displayed in Table~\ref{tab:ampphase}. None of these is significantly different from zero.
The overall integrated CP asymmetry, defined in Eq.~\ref{eqn:intCPasym},
is determined to be $A_{CP}=(-0.05\pm0.57(stat)\pm0.54(syst))\%$, and the systematic uncertainties are shown in Table~\ref{tab:CDFsyst}.
\begin{table}[!htbp]
  \begin{center}
    \begin{tabular}{lcc}
    \hline \hline
    Resonance           &    Amplitude $b$          &  Phase $\phi[\degree]$     \\ \hline
    $K^*(892)^\pm$      & $+0.004\pm0.004\pm0.011$  &   $-0.8\pm1.4\pm1.3$      \\
    $K^*_0(1430)^\pm$   & $+0.044\pm0.028\pm0.041$  &   $-1.8\pm1.7\pm2.2$      \\
    $K^*_2(1430)^\pm$   & $+0.018\pm0.024\pm0.023$  &   $-1.1\pm1.8\pm1.1$      \\
    $K^*(1410)^\pm$     & $-0.010\pm0.037\pm0.021$  &   $-1.6\pm1.9\pm2.2$      \\
    $\rho(770)$         & $-0.003\pm0.006\pm0.008$  &   $-0.5\pm1.5\pm1.4$      \\
    $\omega(782)$       & $-0.003\pm0.002\pm0.000$  &   $-1.8\pm2.2\pm1.4$      \\
    $f_0(980)$          & $-0.001\pm0.005\pm0.004$  &   $-0.1\pm1.3\pm1.1$      \\
    $f_2(1270)$         & $-0.035\pm0.037\pm0.013$  &   $-2.0\pm1.9\pm2.1$      \\
    $f_0(1370)$         & $-0.002\pm0.008\pm0.021$  &   $-0.1\pm1.7\pm2.8$      \\
    $\rho(1450)$        & $-0.016\pm0.022\pm0.135$  &   $-1.7\pm1.7\pm3.9$      \\
    $f_0(600)$          & $-0.012\pm0.017\pm0.025$  &   $-0.3\pm1.5\pm1.4$      \\
    $\sigma_2$          & $-0.011\pm0.012\pm0.004$  &   $-0.2\pm2.9\pm1.1$      \\
    $K^*(892)^\pm$(DCS)     & $+0.001\pm0.005\pm0.002$  &   $-3.8\pm2.3\pm1.2$      \\
    $K^*_0(1430)^\pm$(DCS)  & $+0.022\pm0.024\pm0.035$  &   $-3.3\pm4.0\pm3.9$      \\
    $K^*_2(1430)^\pm$(DCS)  & $-0.018\pm0.029\pm0.017$  &   $+4.2\pm5.3\pm3.0$      \\
    \hline \hline
    \end{tabular}
    \caption{Results of the simultaneous $\DDbar$ Dalitz-plot fit for the CP-violation amplitudes, $b$ and phase, $\phi$. The first uncertainties are statistical and the second systematic.}
    \label{tab:ampphase}
  \end{center}
\end{table}
\begin{table}[!htbp]
  \begin{center}
    \begin{tabular}{lc}
    \hline \hline
    Effect		&  Uncertainty on $A_{CP}$[$10^{-2}$]  \\ \hline
    Efficiency	&  0.36		\\
    Background 	&  0.09  	\\
    Fit model	&  0.37		\\
    Trigger		&  0.05		\\
    Form factors&  0.10 \\ \hline
    Total systematic	& 0.54  \\ \hline
    Statistical	&  0.57	\\ \hline \hline
    \end{tabular}
    \caption{Uncertainties on the overall integrated CP asymmetry.}
    \label{tab:CDFsyst}
  \end{center}
\end{table}

\section{Updated measurement in $D^0\rightarrow K_S\pi^+\pi^-$ at Belle}
We report the updated measurement of $D^0-\overline{D}^0$ mixing
by time-dependent Dalitz analysis method using the $921~fb^{-1}$ of $\Upsilon$(4S) and
$\Upsilon$(5S) full data collected by Belle detector at
KEKB asymmetric-energy $e^+e^-$ collider.

The decay chain $D^{*+}\rightarrow D^0\pi_s^+$, $D^0\rightarrow K_S^0\pi^+\pi^-$,
which is reconstructed from $c\bar{c}$ process, is used to distinguish between $D^0$
and $\overline{D}^0$ with the charge of the low-momentum pion $\pi_s$ and to reduce the background.
The $D^0$ decay time $t$ and its uncertainty $\sigma_t$
are obtained by projecting the flight length between $D^0$ decay and production vertices to momentum direction and then transforming it
to the center-of-mass system(CMS). To suppress the combinatorial background and the events from B decays,
we require $D^{*+}$ momentum in CMS to be greater than $2.5$ GeV/c and $3.1$ GeV/c for $\Upsilon$(4S) and
$\Upsilon$(5S) data respectively.

\begin{figure}[!htpb]
  \begin{center}
    \includegraphics[width=0.25\paperwidth]{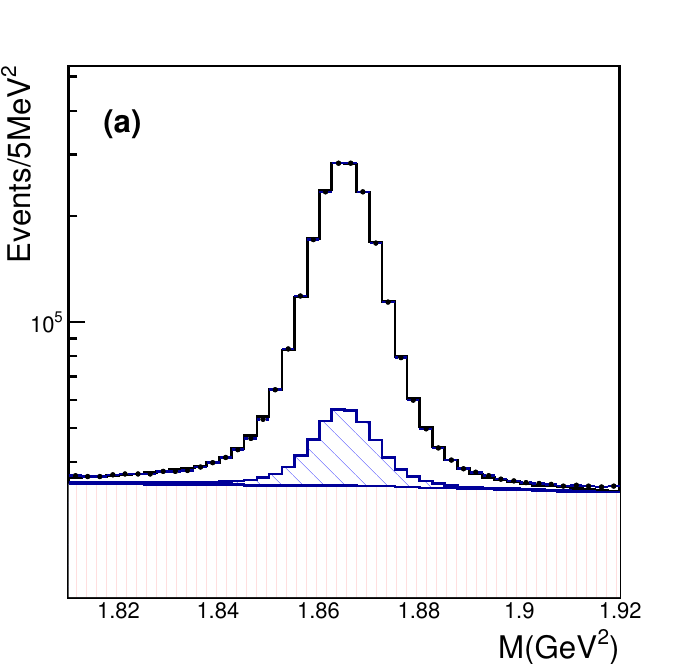}%
    \includegraphics[width=0.25\paperwidth]{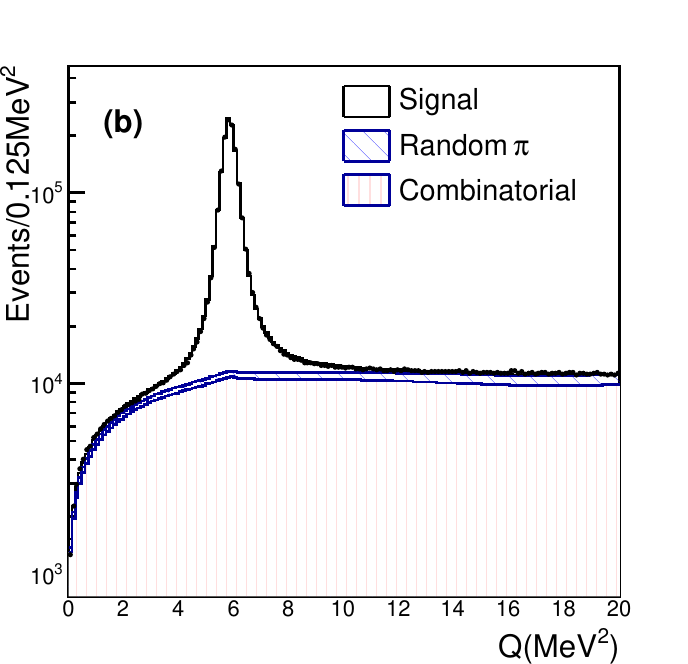}
    \vskip-10pt
    \caption{The projections of $M$ and $Q$ for MC data in region $1.81<M<1.92~GeV/c^2$ and $0<Q<20~MeV/c^2$, including signal, random $\pi$ background and combinatorial background.}
    \label{fig:MQdistr}
  \end{center}
\end{figure}

Two observable are used to determined the yield of signal and backgrounds:
the invariant mass of $D^0$ daughter particles:
$M=m_{K_S^0\pi^+\pi^-}$ and the energy released from $D^{*+}$ decay:
$Q=m_{K_S^0\pi^+\pi^-\pi_s}-m_{K_S^0\pi^+\pi^-}-m_{\pi_s}$. The $M$ and $Q$ distributions of
selected candidates are shown in Figure~\ref{fig:MQdistr}.
The Dalitz distribution of $D^0\rightarrow K_S^0\pi^+\pi^-$ are expressed as a sum of quasi-two-body amplitudes.
Different modes are used to describe the decay amplitude:
Breit-Weigner model for the P- and D-wave with twelve intermediate resonances,
K-matrix model for the $\pi\pi$ S-wave~\cite{BaBarCKMgamma} and LASS model for $K_S^0\pi$ S-wave~\cite{BaBar2010}. The final Dalitz plot
parameterize is optimized according likelihood and $\chi^2$ test.

\begin{figure}[!hbtp]
  \begin{center}
    \includegraphics[width=0.25\paperwidth]{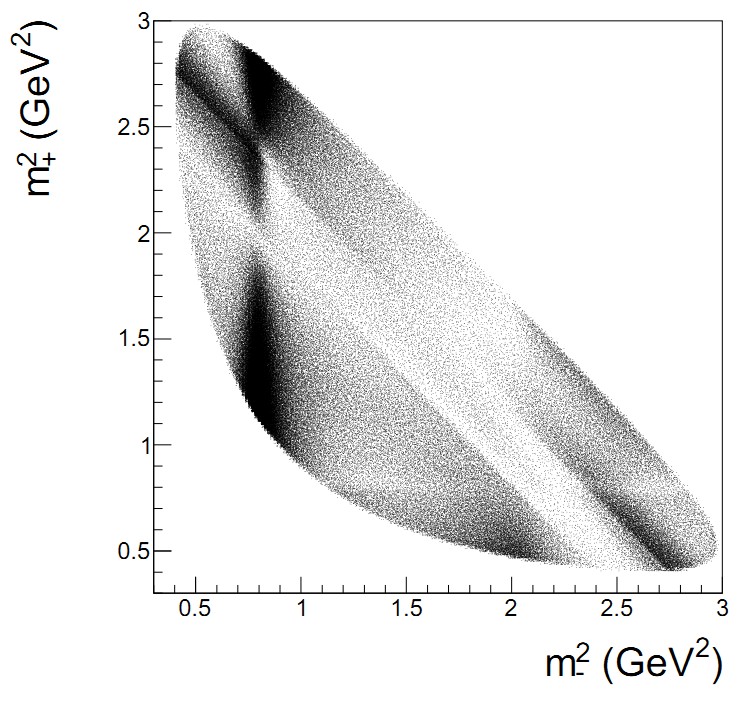}%
    \includegraphics[width=0.25\paperwidth]{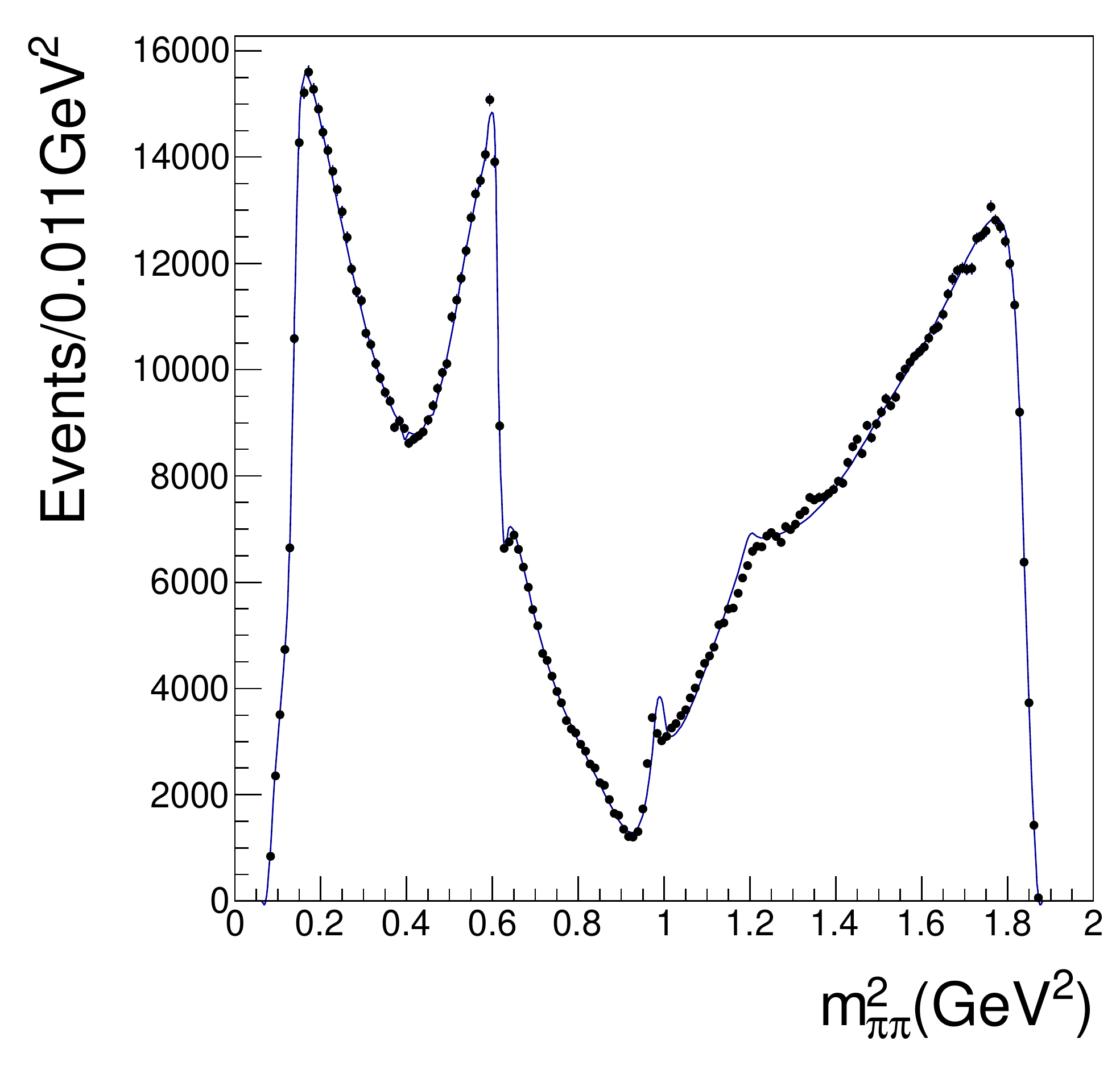}\break
    \includegraphics[width=0.25\paperwidth]{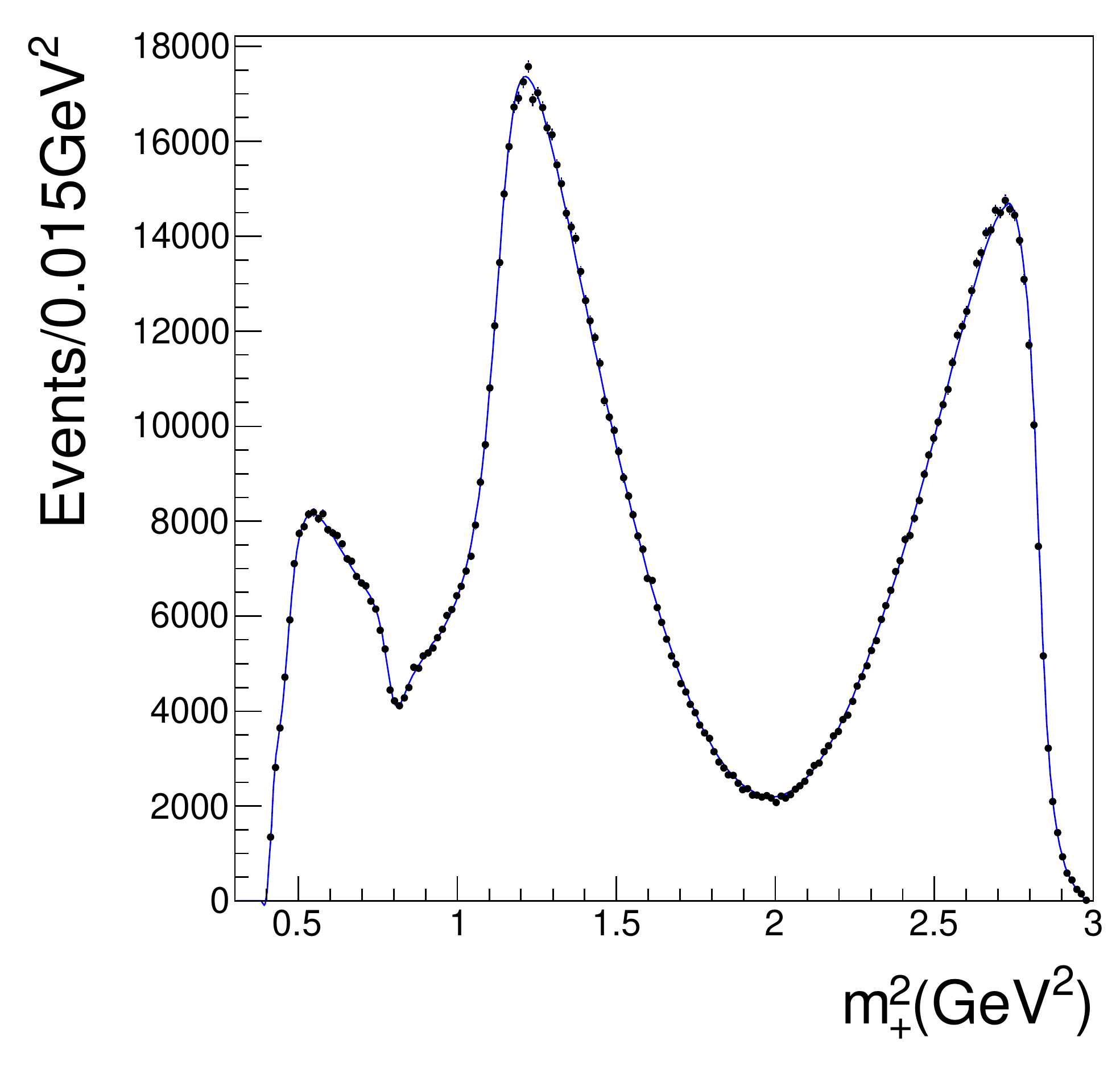}%
    \includegraphics[width=0.25\paperwidth]{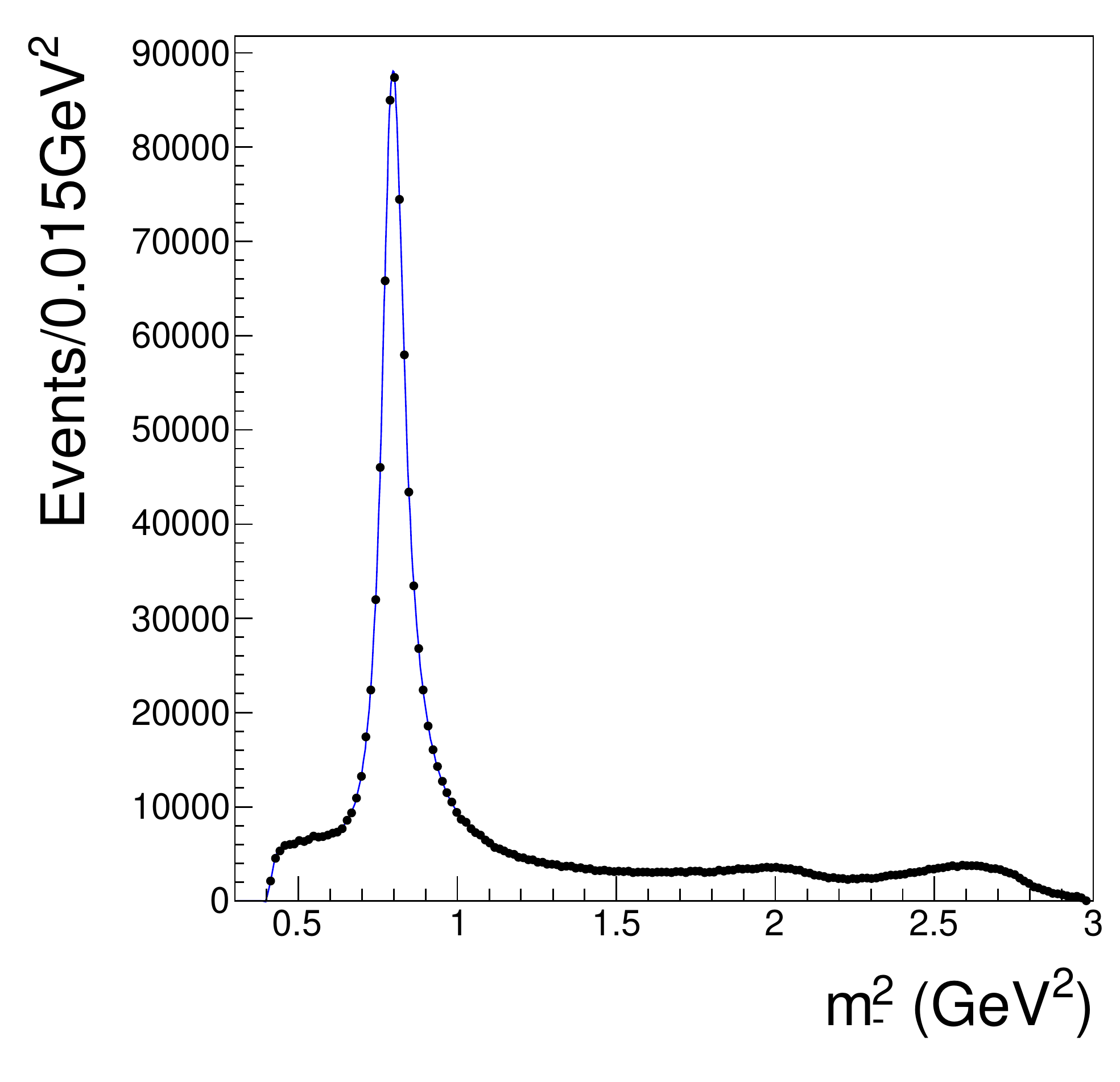}
    \vskip-10pt
    \caption{Dalitz plot distribution and Dalitz variables $m^2_+$, $m^2_-$ and $m^2_{\pi\pi}$ projections for experimental data with 2-dimensional $\chi^2$ test over the Dalitz plot plane: $\chi^2/ndf=1.207$ for $14264-42$ degrees of freedom.}
    \label{fig:Dalitzplot}
  \end{center}
\end{figure}

\begin{figure}[!hbtp]
  \begin{center}
    \includegraphics[width=0.333\paperwidth]{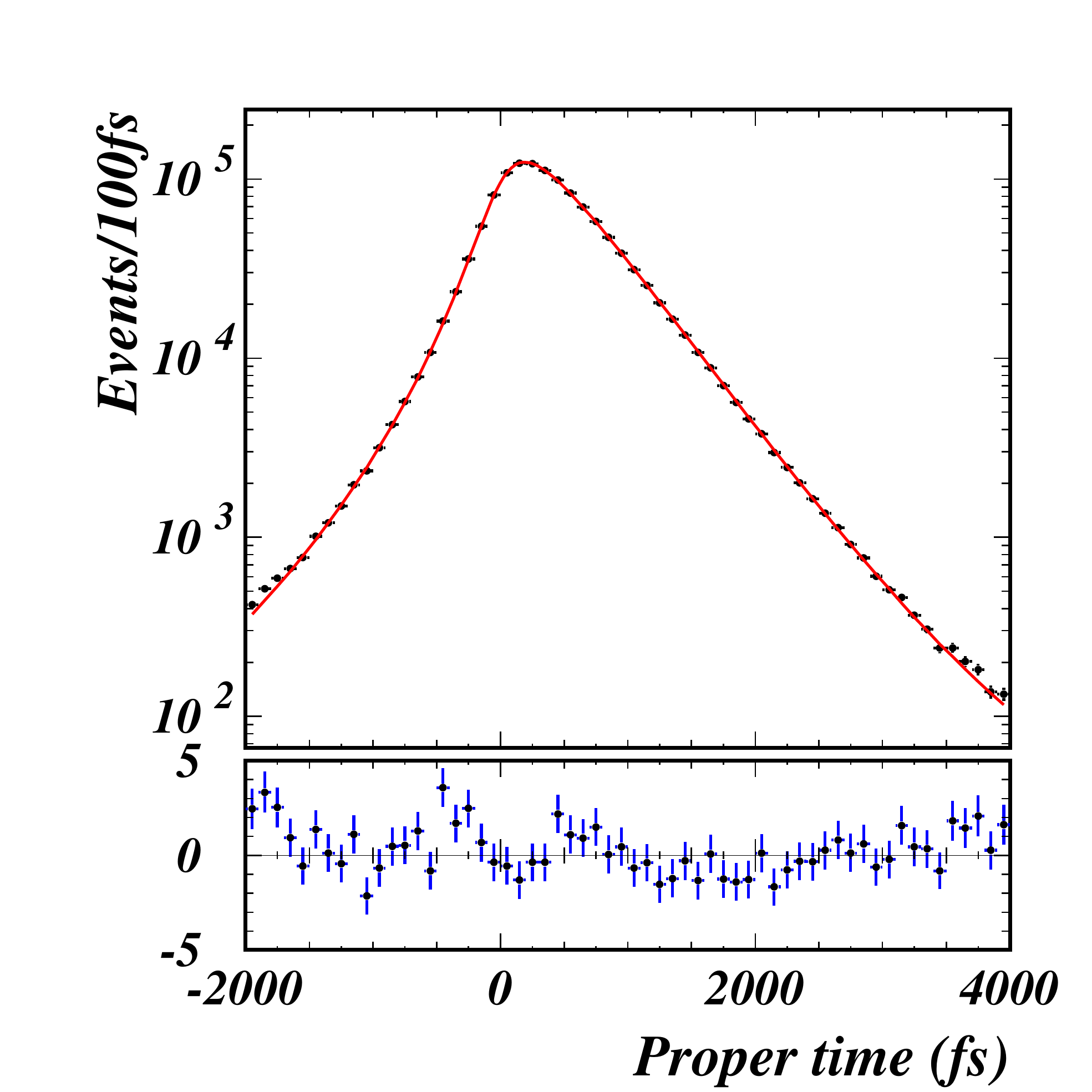}
    \vskip-10pt
    \caption{The proper time distribution for events in the signal region with $-2000<t<4000(fs)$(point) and fit projection for mixing fit(curve).}
    \label{fig:propertime}
  \end{center}
\end{figure}

We performe an unbinned maximum likelihood fit to extract the mixing parameters $x$ and $y$. The distribution of
combinatorial background is determined in $M$ sideband($30<|M-m_{D^0}|<50~MeV/c^2$).
Meanwhile the random $\pi^+$ background is mixture of true and
mistagged $D^0$ with the fraction determined from $Q$ sideband($3<|Q-5.85|<14.15~MeV/c^2$).
In the final Dalitz fit with data sample, see Figure~\ref{fig:Dalitzplot}, we extract the mixing parameters
$x=(0.56\pm0.19^{+0.03+0.06}_{-0.09-0.09})\%$ and $y=(0.30\pm0.15^{+0.04+0.03}_{-0.05-0.06})\%$, where the errors are statistical,
experimental and model uncertainties respectively, see Table~\ref{tab:sysun},
and the $D^0$ mean lifetime $\tau=(410.3\pm0.4)fs$, see Figure~\ref{fig:propertime}, which is consistent with the world average.

\begin{table}
\begin{center}
\begin{tabular}{lcc}
  \hline \hline
  Source of systematic uncertainty   & ($\Delta x$)($\times10^{-4}$) & ($\Delta y$)($\times10^{-4}$)  \\ \hline
  Best Candidate selection              & $+1.05$       &   $+1.87$     \\
  Signal and backgrounds yields         & $\pm0.30$     &   $\pm0.27$  \\
  Wrong tagged events' fraction         & $-0.67$       &   $-0.45$  \\
  Time resolution of signal             & $-1.39$       &   $-0.92$  \\
  Efficiency                            & $-1.13$       &   $-2.09$  \\
  Combinatorial's PDF                   & $^{+1.90}_{-4.82}$    & $^{+2.28}_{-3.88}$ \\
  $K^*(892)$ DCS/CF reduced by $5\%$    & $-7.28$       &   $+2.29$  \\
  $K^*(1430)$ DCS/CF reduced by $5\%$   & $+1.71$       &   $-0.67$  \\  \hline
  Total uncertainty(experimental sys.)                      & $^{+2.78}_{-8.94}$    & $^{+3.74}_{-4.58}$ \\
  \hline \hline
  Resonances' $M$ and $\Gamma$ error    & $\pm1.40$     &   $\pm1.21$       \\
  Remove $K^*(1680)^+$                  & $-1.78$       &   $-3.02$         \\
  Remove $K^*(1410)^{\pm}$              & $-1.16$       &   $-3.62$         \\
  Remove $\rho(1450)$                   & $+2.13$       &   $+0.30$         \\
  Form factors                          & $+4.05$       &   $+2.35$         \\
  $\Gamma(q^2)=$constant                & $+3.33$       &   $-1.61$         \\
  Angular dependence                    & $-8.46$       &   $-3.86$         \\
  K-matrix formalism                    & $-2.16$       &   $+1.79$         \\ \hline
  Total uncertainty(model sys.)                     & $^{+5.83}_{-9.09}$    & $^{+3.21}_{-6.42}$      \\
  \hline \hline
\end{tabular}
\caption{The source of two kinds of systematic uncertainty.}
\label{tab:sysun}
\end{center}
\end{table}
We also search for CP violation in $D^0/\overline{D}^0\rightarrow K_S^0\pi^+\pi^-$ decays. We obtain identical mixing parameters as fit
result without CP violation and these CP violation parameters $|q/p|=0.90^{+0.16+0.05+0.06}_{-0.15-0.04-0.05}$
and $\arg(q/p)(^o)=-6\pm11^{+3+3}_{-3-4}$ which show no hint for indirect CP violation.
\Acknowledgements
I am grateful to T. Peng, Z.P. Zhang, A. Zupanc, J. Brodzicka, M. Nakao for advices
and collaboration for the contribution for KEKB in achieving the highest luminosity
and most stable machine condition.

\end{document}